\documentclass[a4paper,11pt]{article}
\usepackage{pos}
\usepackage{amsmath}

\usepackage{amssymb}
\usepackage{bm}
\usepackage{physics}
\usepackage{caption}
\usepackage{subcaption}

\title{A Bayesian Pixel Based Approach for Model Independent TMD Reconstruction}
\ShortTitle{Bayesian Pixel Based TMD Reconstruction}

\author*{Marco Zaccheddu}

\affiliation{Theory Center, Jefferson Lab, 12000 Jefferson Avenue, Newport News, Virginia 23606, USA}

\emailAdd{zacch@jlab.org}

\abstract{We introduce a nonparametric pixel-based framework for the Bayesian inference and imaging of transverse momentum dependent (TMD) parton distributions. The methodology integrates TMD evolution within the Collins-Soper-Sterman formalism in a differentiable framework, and leverages generative AI through a hybrid normalizing flow-driven Metropolis-Hastings algorithm for efficient posterior sampling. The framework is validated through multi-scale closure tests of increasing complexity. Using singular value decomposition, we characterize the existence of {\it null TMDs}, functional components that remain unconstrained by observables, and demonstrate how multi-scale data break these degeneracies, enabling 3D partonic imaging.}

\notes{\vspace{0.8em}\noindent\hspace{0.28cm}{\small\itshape Report No.~JLAB-THY-26-4937}}

\FullConference{The 33rd International Workshop on Deep Inelastic Scattering and Related Subjects (DIS2026)\\
4 - 8 May 2026\\
Bologna, Italy\\}

\begin{document}
\maketitle

\section{Introduction}
\label{sec:intro}

The extraction of transverse momentum dependent (TMD) parton distribution functions (PDFs) and fragmentation functions (FFs) is governed by factorization theorems within the perturbative QCD framework~\cite{Collins:1984kg, Collins:2011zzd}. In processes such as semi-inclusive deep-inelastic scattering (SIDIS) and Drell-Yan production, the cross sections are expressed as convolutions of two TMD distributions. Within the Collins-Soper-Sterman (CSS) formalism~\cite{Collins:1984kg, Collins:2011zzd}, this convolution is naturally formulated in impact parameter ($b_T$) space, related to transverse momentum ($k_T$) space via a Bessel transform:
\begin{equation}
	W(k_T) = \frac{1}{2\pi} \int_{0}^{\infty} \dd b_{T}\, b_{T}\, J_0(b_{T} \,k_{T})\, \widetilde{W}(b_{T}) \, .
\label{eq:fredholm}
\end{equation}
The determination of $\widetilde{W}(b_{T})$ from experimental data constitutes a Fredholm integral equation of the first kind~\cite{Tikhonov:1977, Aster:2018}, which is intrinsically ill-posed: infinitely many configurations in $b_T$ space can reproduce the same observables within uncertainties. State-of-the-art TMD extractions~\cite{Scimemi:2019cmh, Moos:2023yma, Bacchetta:2022awv, Barry:2023qqh} typically address this by assuming rigid analytical functional forms or, more recently, neural network parametrizations~\cite{Bacchetta:2025nn, Fernando:2025xzv}, alongside theoretical efforts to analytically continue resummation into the deep infrared~\cite{Simonelli:2025kga}. While successful, these approaches remain insensitive to functional components residing in the null space of the integral transform, potentially underestimating the true uncertainty of the 3D partonic map.

In this work~\cite{Zaccheddu:2026pyy}, we introduce a nonparametric, pixel-based Bayesian framework that enables the identification and quantification of such {\it null TMDs}, analogous to the shadow distributions encountered in the study of generalized parton distributions~\cite{Bertone:2021, Moffat:2023}. By replacing rigid functional forms with a discrete grid of nodal values, we allow the data to dictate the shape of the distribution with minimal {\it a priori} bias. The high-dimensional posterior is sampled using a hybrid normalizing flow-driven Metropolis-Hastings (NF-MH) algorithm~\cite{Dinh:2016}. The methodology is validated through three closure tests of increasing complexity: a baseline Gaussian model, the $u$-quark TMD PDF, and the unpolarized $F_{UU,T}$ structure function.

\section{Methodological Framework}
\label{sec:methodology}

\subsection{Discretization and Forward Mapping}

The unknown distribution $\tilde{f}(b_T)$ is approximated within a finite interval using a basis of $N$ local interpolators $h_i(b_T)$ defined on a power-law grid of nodes~\cite{Freese:2024ypk}:
\begin{equation}
\tilde{f}(b_T) \approx \sum_{i=1}^{N} \tilde{f}_i\, h_i(b_T)\,,
\label{eq:bt_interpolation}
\end{equation}
where the nodal values $\tilde{f}_i$ are referred to as {\it pixels}. Substituting into the Bessel transform yields a discrete matrix equation $\boldsymbol{f} = \mathcal{M}\, \tilde{\boldsymbol{f}}$, where the kernel matrix elements are:
\begin{equation}
\mathcal{M}_{ji} = \frac{1}{2\pi} \int_{b_{T\min}}^{b_{T\max}} \dd b_{T}\, b_{T}\, J_0(b_{T} k_{T j})\, h_i(b_{T}).
\label{eq:kernel_matrix}
\end{equation}
To handle scale evolution, the pixels vector at the observation scale $\mu^2$ is decomposed as the Hadamard product of the reference distribution and the evolution operator: $\tilde{\boldsymbol{f}}(\mu^2) = \tilde{\boldsymbol{f}}(\mu_0^2) \odot \mathbf{U}(\mu^2)$, ensuring the inference targets the intrinsic distribution at a universal reference scale $\mu_0^2$.

\subsection{Bayesian Inference and NF-MH Sampling}

The posterior distribution $P(\tilde{\boldsymbol{f}} | \boldsymbol{f}) \propto \mathcal{L}(\boldsymbol{f} | \tilde{\boldsymbol{f}})\, \pi(\tilde{\boldsymbol{f}})$ is sampled using a hybrid NF-MH strategy. The likelihood is Gaussian, with a $\chi^2$ cost function comparing the forward projection $\mathcal{M}\tilde{\boldsymbol{f}}$ to the data. The prior employs Tikhonov regularization up to second order, penalizing unphysical amplitudes, sharp gradients, and rapid oscillations.
A normalizing flow (NF), composed of 12 masked affine coupling layers, is first trained to approximate the posterior by minimizing the KL divergence. The trained NF then serves as a global proposal distribution for an independence-sampler Metropolis-Hastings chain, which corrects residual biases via the acceptance probability:
\begin{equation}
\alpha = \min \!\left( 1, \frac{P(\tilde{\boldsymbol{f}}_{\text{new}} | \boldsymbol{f}) }{P(\tilde{\boldsymbol{f}}_{\text{old}} | \boldsymbol{f}) } \frac{q^{\text{NF}}(\tilde{\boldsymbol{f}}_{\text{old}})}{q^{\text{NF}}(\tilde{\boldsymbol{f}}_{\text{new}})} \right)\,.
\label{eq:mh_acceptance}
\end{equation}
%

\section{Gaussian Case Study: Results and SVD Analysis}
\label{sec:gaussian}

\begin{figure}[t]
	\centering
	\includegraphics[width=0.86\textwidth]{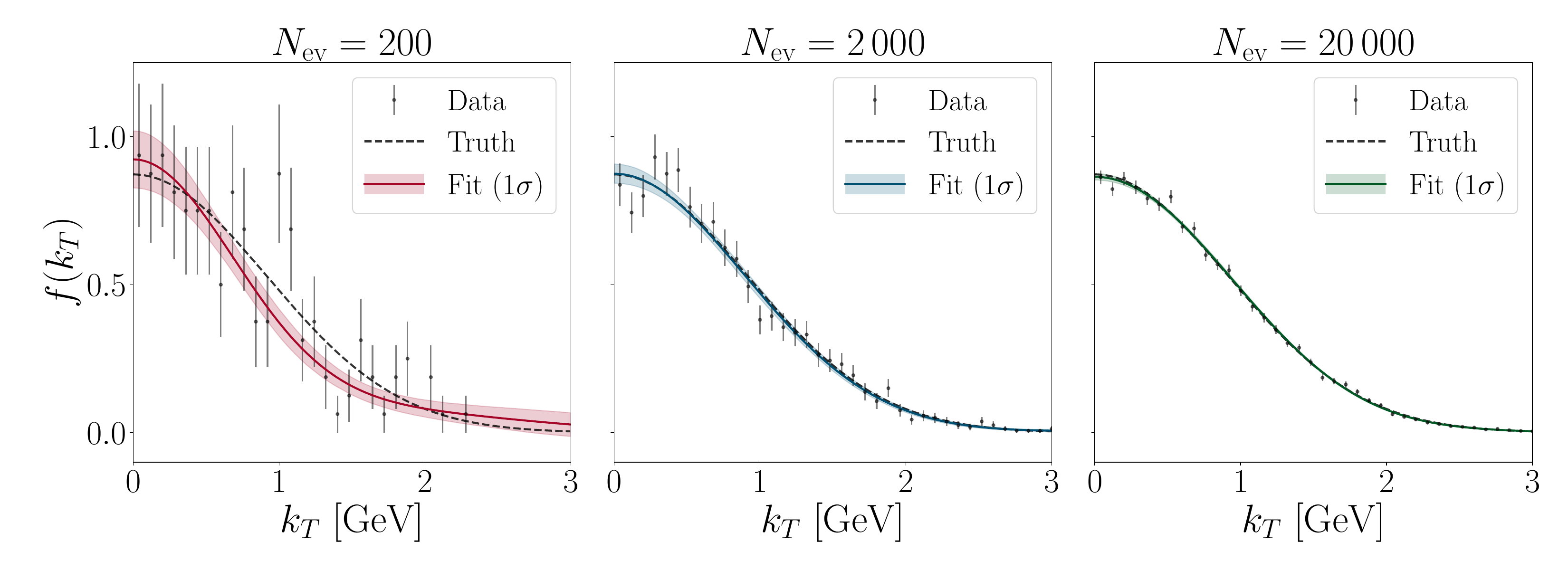}
	\caption{Comparison between pseudo-data (markers with error bars) and the Bayesian fit (solid curves and 1$\sigma$ shaded uncertainty bands) in $k_T$ space for $N_{\text{ev}} = 200$, $2\,000$, and $20\,000$.}
	\label{fig:scaling_fits_kt}
\end{figure}

\begin{figure}[t]
	\centering
	\includegraphics[width=0.86\textwidth]{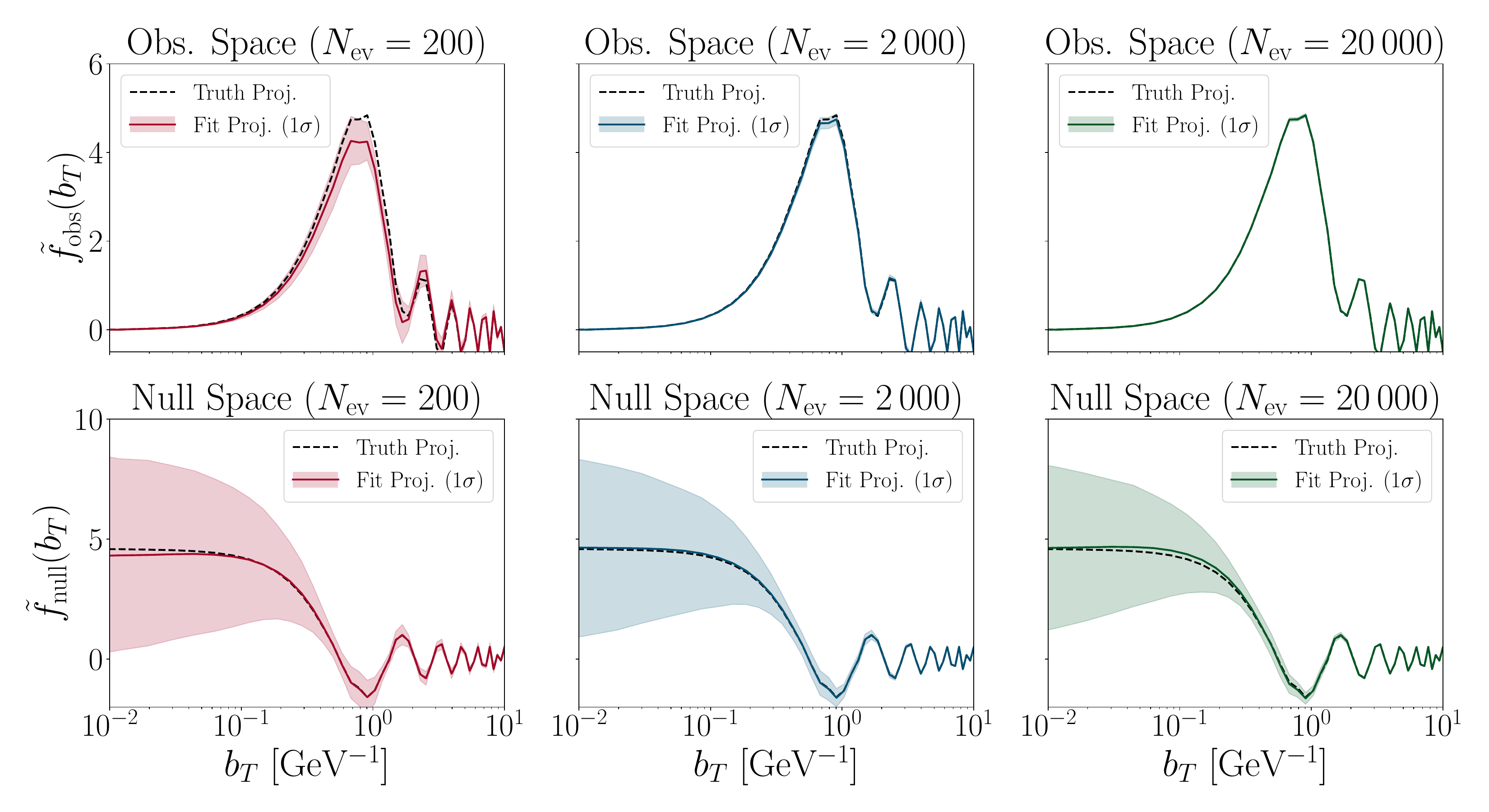}
	\caption{Decomposition of the fit uncertainty into observable (top row) and null (bottom row) subspaces across statistical regimes. The statistical gain is only visible in the observable space, while the null space remains unchanged, establishing an irreducible precision floor.}
	\label{fig:projection_scaling}
\end{figure}

The framework is first validated on a Gaussian model where the ground truth is known analytically. We perform both a single-scale ($\mu^2 = 2$~GeV$^2$) and a multi-scale ($\mu^2 \in \{2, 5, 20, 50\}$~GeV$^2$) analysis across different statistical regimes ($N_{\text{ev}} = 200$, $2\,000$, $20\,000$).
In the single-scale analysis, the reconstructed distributions in $k_T$ space achieve $\chi^2/N_{\text{pts}} < 1$ across all regimes, as shown in Fig.~\ref{fig:scaling_fits_kt}. In $b_T$ space, however, a ``precision floor'' emerges: the uncertainty at small $b_T$ does not shrink with increasing statistics, in sharp contrast to the ideal $1/\sqrt{N_{\text{ev}}}$ scaling observed in momentum space.
The origin of this behavior is revealed by the singular value decomposition (SVD) of the kernel matrix $\mathcal{M} = U \Sigma V^T$. The rapid decay of the singular values defines an effective rank $p$, partitioning the model space into an {\it observable} subspace $V_{\text{obs}}$ (spanned by $\{\mathbf{v}_1, \dots, \mathbf{v}_p\}$) and a {\it null} subspace $V_{\text{null}}$ (spanned by $\{\mathbf{v}_{p+1}, \dots, \mathbf{v}_N\}$). Since $\mathcal{M}\,\mathbf{v}_i \approx 0$ for modes $i > p$, the likelihood is insensitive to null-space components, and the posterior for $\tilde{\boldsymbol{f}}_{\text{null}}$ is governed by the prior. This formally defines the {\it null TMDs}: functional modes that remain invisible to observables regardless of the available statistics.
The decomposition of the posterior into observable and null components (Fig.~\ref{fig:projection_scaling}) confirms that the observable uncertainty scales as $1/\sqrt{N_{\text{ev}}}$, while the null-space uncertainty remains invariant, establishing an irreducible precision floor.
In the multi-scale analysis, including data at four energy scales substantially improves the reconstruction. The evolution-induced broadening of the $k_T$ spectrum at higher energies provides the high-frequency information needed to resolve the small-$b_T$ structure. The observable subspace expands, and the evolution parameter $g_{\text{evo}}$ is simultaneously extracted with a precision that improves by nearly a factor of 20 between $N_{\text{ev}} = 200$ and $20\,000$ (Table~\ref{tab:gauss_multi}).

\begin{table}[t]
	\caption{Global fit diagnostics for the multi-scale Gaussian case. The ground truth is $g_{\text{evo}} = 1.2$~GeV$^2$.}
	\centering
	\begin{tabular}{rclcc}
		\hline
		$N_{\text{ev}}$~ & accept. rate & $g_{\text{evo}} \, [{\rm GeV}^2]$  & $N_{\text{pts}}$ &  $\chi^2/N_{\text{pts}}$ \\
		\hline
		$200$     & 59\% & ~1.2(3)    & 148 & 0.99 \\
		$2\,000$  & 82\% & ~1.10(5)   & 207 & 1.05 \\
		$20\,000$ & 82\% & ~1.209(15) & 248 & 0.99 \\
		\hline
	\end{tabular}
	\label{tab:gauss_multi}
\end{table}

\section{TMD PDF Extraction}
\label{sec:tmd}

The framework is then extended to the full TMD PDF for the $u$-quark flavor. The $b_T$-space distribution follows the CSS formalism:
\begin{equation}
\begin{split}
\tilde{f}_{f/P}(x, b_T; Q) &= \sum_j \int_x^1 \frac{\dd\hat{x}}{\hat{x}}\, C_{f/j}\big(x/\hat{x}, b_*; \mu^2_b, \mu_b, g(\mu_b)\big)\, f_{j/P}(\hat{x}, \mu_b) \\
&\quad \times \exp\! \big[ S_{\rm pert}(b_T;\mu_b, Q) \big] M_f(x,b_T) \exp\! \Big[\! - g_K(b_T) \ln \frac{Q}{Q_{0}} \Big]\, ,
\end{split}
\label{eq:PDF_func}
\end{equation}
where $S_{\rm pert}$ is the perturbative Sudakov factor at NLO, and $M_f$, $g_K(b_T)$ encode the nonperturbative structure and evolution, respectively. The intrinsic distribution is modeled as $M_f(b_T) = R(b_T) + \tilde{p}(b_T)[1 - R(b_T)]$, where $R(b_T)$ is a logistic mask enforcing the perturbative limit at small $b_T$, and $\tilde{p}(b_T)$ is the pixel-based unknown. Collinear inputs are taken from the JAM20 set~\cite{Moffat:2021mxy}.

At a single scale ($Q = 2$~GeV), the algorithm achieves $\chi^2/N_{\text{pts}} < 1$ with acceptance rates above 69\%. The resolution analysis confirms the pattern observed in the Gaussian case: the null space dominates at small $b_T$, with the resolution indices near zero across most of the spectrum.

\begin{figure}[t]
	\centering
	\includegraphics[width=0.86\textwidth]{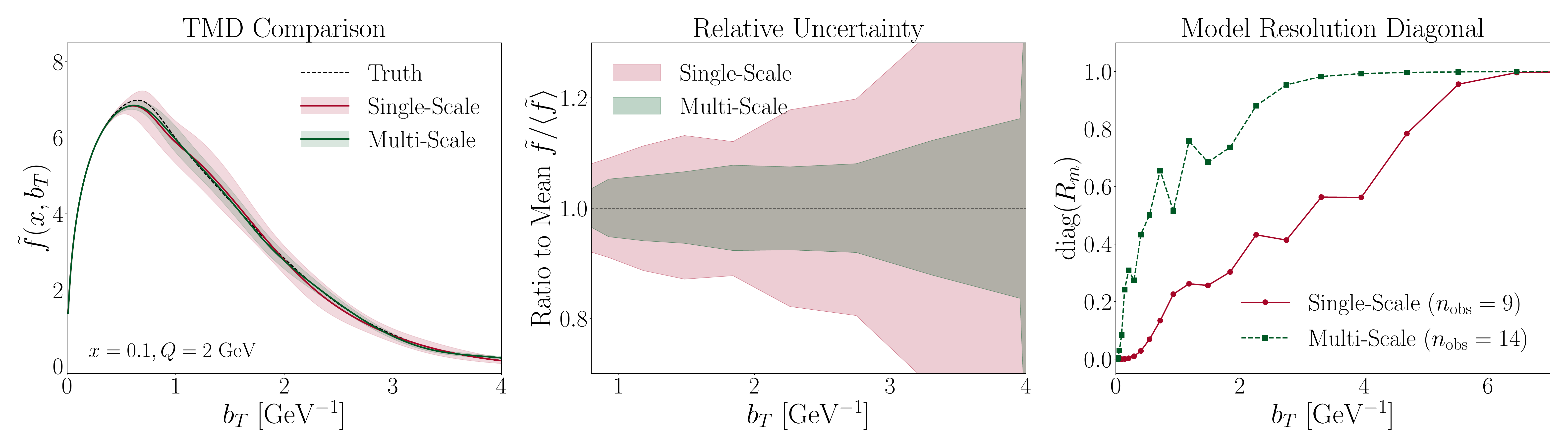}
	\caption{Comparison between single-scale ($Q=2$~GeV, red dashed curves and 1$\sigma$ bands) and multi-scale (blue solid curves and 1$\sigma$ bands) TMD PDF extractions for $N_{\text{ev}} = 10^5$. Left: reconstructed $b_T$-space distributions compared to ground truth (black dot-dashed curve). Center: relative uncertainties. Right: resolution indices.}
	\label{fig:single_vs_multi}
\end{figure}

In the multi-scale scenario ($Q \in \{2, 5, 7, 10\}$~GeV), the simultaneous extraction of $M_f(b_T)$ and $g_2$ yields a substantially improved reconstruction (Fig.~\ref{fig:single_vs_multi}). The extracted evolution parameter converges to the ground truth ($g_2 = 0.28$~GeV$^2$) with uncertainties shrinking from 0.02 to 0.003 as $N_{\text{ev}}$ increases from $10^3$ to $10^5$. The resolution indices confirm the expanded observable subspace.

\section{\texorpdfstring{$F_{UU,T}$}{F\_UU,T} Structure Function}
\label{sec:fuu}

The most challenging test involves the extraction from the unpolarized structure function $F_{UU,T}$, which requires deconvolving the PDF from the TMD FF within the SIDIS factorization framework~\cite{Bacchetta:2006tn}:
\begin{equation}
F_{UU,T}(x, z, q_T, Q^2) = x \sum_a e_a^2 \frac{1}{2\pi} \int_0^{\infty} \dd b_T \, b_T\, J_0(q_T b_T)\, \tilde{f}_1^a(x, b_T; Q)\, \widetilde{D}_1^a(z, b_T; Q) \, .
\label{eq:fuu}
\end{equation}
The FF introduces an {\it effective kernel} $\mathcal{M}_{\text{eff}}$ that modulates the sensitivity to the $b_T$ space. At large impact parameters, the FF vanishes, masking the large-$b_T$ structure and leading to a sharp loss of spatial resolution beyond $b_T \approx 6~\text{GeV}^{-1}$.
Despite this additional complexity, the NF-MH algorithm successfully deconvolves the PDF signal across all statistical regimes.

\begin{figure}[t]
	\centering
	\includegraphics[width=0.84\textwidth]{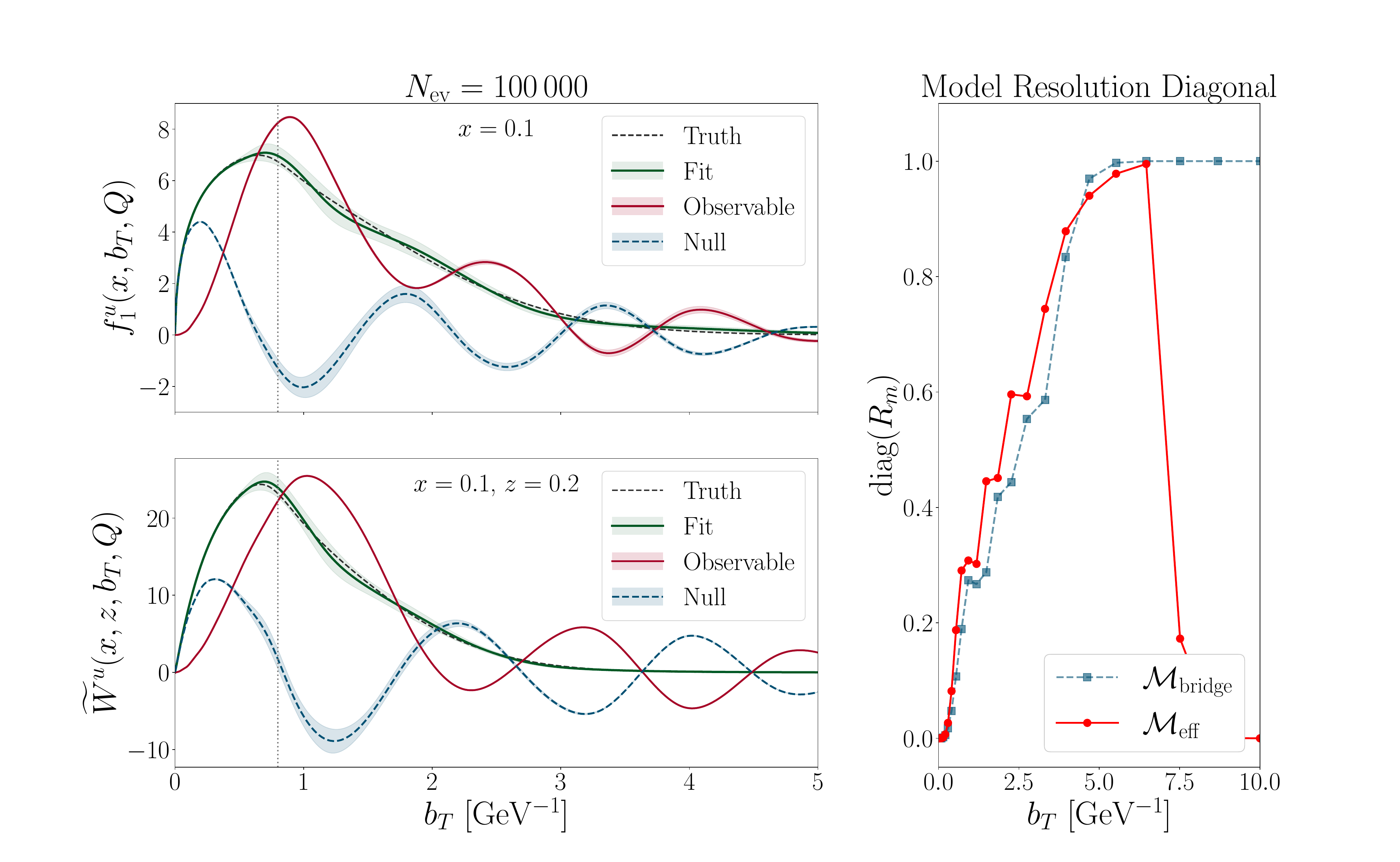}
	\caption{Resolution analysis for the $F_{UU,T}$ case at $N_{\text{ev}} = 10^5$. (Top-Left) Reconstructed TMD PDF with projections onto the observable space (red solid curve) and null space (blue dashed curve) using $\mathcal{M}_{\text{eff}}$. (Bottom-Left) Reconstructed $\widetilde{W}$ with projections using $\mathcal{M}_{\text{bridge}}$. (Right) Resolution indices for $\mathcal{M}_{\text{eff}}$ (red solid curve) and $\mathcal{M}_{\text{bridge}}$ (blue dashed curve).}
	\label{fig:fuu_resolution}
\end{figure}

The resolution analysis at $10^5$ events (Fig.~\ref{fig:fuu_resolution}) reveals how information is partitioned between the observable and null spaces in the presence of the fragmentation convolution. The effective kernel provides higher resolution than the pure bridge matrix up to $b_T \approx 4~\text{GeV}^{-1}$, but suffers a sharp drop in resolution at larger distances where the FF vanishes. This confirms that the FF sets a physical limit on the spatial resolution of the 3D partonic map.
In the multi-scale fit ($Q \in \{2, 5, 7, 10\}$~GeV), the evolution parameter $g_2$ is recovered within uncertainties (Table~\ref{tab:fuu_multi}), and the reconstructed TMD PDF agrees with the ground truth across the full $b_T$ range. The multi-energy lever arm suppresses the null-space dominance at small and intermediate distances, pushing the precision floor toward shorter distances.

\begin{table}[t]
	\caption{Global fit diagnostics for the multi-scale $F_{UU,T}$ case. The ground truth is $g_2 = 0.28$~GeV$^2$.}
	\centering
	\begin{tabular}{lclcc}
		\hline
		$N_{\text{ev}}$ & accept. rate & $g_2\,[{\rm GeV^2}]$  & $N_{\text{pts}}$ & $\chi^2/N_{\text{pts}}$ \\
		\hline
		$10^3$   & 84\% & ~$0.22(5)$ & 251 & 1.02 \\
		$10^4$   & 85\% & ~$0.28(2)$ & 276 & 0.94 \\
		$10^5$   & 68\% & ~$0.272(5)$ & 284 & 1.07 \\
		\hline
	\end{tabular}
	\label{tab:fuu_multi}
\end{table}

\section{Conclusions}
\label{sec:conclusions}

We have introduced a nonparametric pixel-based framework for TMD imaging that integrates Bayesian inference, generative AI (normalizing flows), and SVD diagnostics. The SVD analysis provides the first formal characterization of {\it null TMDs}, functional modes that remain unconstrained by single-scale data. We demonstrated that multi-scale measurements act as a kinematic lever arm to break these degeneracies, expanding the observable subspace and enabling 3D partonic imaging. The framework has been validated through closure tests of increasing complexity, from basic Gaussian models to the full $F_{UU,T}$ structure function, establishing a robust foundation for future applications to experimental data from Jefferson Lab and the Electron-Ion Collider.

\acknowledgments

This material is based upon work supported by the U.S. Department of Energy, Office of Science, Office of Nuclear Physics under Contract No. 89243126CSC000213, and in part under the Laboratory Directed Research and Development (LDRD) Program at Thomas Jefferson National Accelerator Facility for the U.S.~Department of Energy. 

\let\oldthebibliography\thebibliography
\let\endoldthebibliography\endthebibliography
\renewenvironment{thebibliography}[1]{%
  \begin{oldthebibliography}{#1}%
    \small
    \setlength{\itemsep}{1pt plus 0.5pt}%
    \setlength{\parskip}{0pt}%
}{%
  \end{oldthebibliography}%
}

\bibliographystyle{JHEP}
\bibliography{proceedings}

\end{document}